\documentclass[twocolumn,showpacs,amsmath,amssymb,aps,prl,floatfix,nofootinbib,superscriptaddress]{revtex4}
\usepackage{graphicx,graphics,times,color}
\usepackage{bm}
\usepackage{longtable}
\usepackage{color}

\def\be{\begin{equation}}
\def\ee{\end{equation}}
\def\ba{\begin{eqnarray}}
\def\ea{\end{eqnarray}}
\def\la{\langle}
\def\ra{\rangle}

\begin{document}

\title{Entanglement probe of two-impurity Kondo physics in a spin chain}

\author{Abolfazl Bayat}
\affiliation{Institut f\"{u}r Theoretische Physik, Albert-Einstein Allee 11, Universit\"{a}t Ulm, 89069 Ulm, Germany}
\affiliation{Department of Physics and Astronomy, University College
London, Gower St., London WC1E 6BT, United Kingdom}
\author{Sougato Bose}
\affiliation{Department of Physics and Astronomy, University College
London, Gower St., London WC1E 6BT, United Kingdom}

\author{Pasquale Sodano}
\affiliation{Department of Physics, University of Perugia, and INFN, Sezione di Perugia, Via A. Pascoli, 06123, Perugia, Italy}
\affiliation{ International Institute of Physics, Federal University of Rio Grande do Norte, Av. Odilon Gomes de Lima 1722, Capim Macio, Natal-RN 59078-400, Brazil}
\author{Henrik Johannesson}
\affiliation{Department of Physics, University of Gothenburg, SE 412 96 Gothenburg, Sweden}

\begin{abstract}
We propose that real-space properties of the two-impurity Kondo model can be obtained from an effective spin model where two single-impurity Kondo spin chains are joined via an RKKY interaction between the two impurity spins. We then use a DMRG approach, valid in all ranges of parameters, to study its features using two complementary quantum-entanglement measures, the {\em negativity} and the {\em von Neumann entropy}. This non-perturbative approach enables us to uncover the precise dependence of the spatial extent $\xi_K$ of the Kondo screening cloud with the Kondo and RKKY  couplings. Our results reveal an exponential suppression of the Kondo temperature $T_K \sim 1/\xi_K$ with the size of the effective impurity spin in the limit of large ferromagnetic RKKY coupling, a striking  display of ``Kondo resonance narrowing" in the two-impurity Kondo model. We also show how the antiferromagnetic RKKY interaction produces an effective decoupling of the impurities from the bulk already for intermediate strengths of this interaction, and, furthermore, exhibit how the non-Fermi liquid quantum critical point is signaled in the quantum entanglement between various parts of the system.
\end{abstract}

\maketitle

{\em Introduction.-}
The theory of quantum impurities underpins much of the current understanding of correlated electrons. A case in point is the {\em two-impurity Kondo model} (TIKM) \cite{Jayprakash}, with bearing on heavy fermion physics \cite{Jones}, correlation effects in nanostructures \cite{Bork}, spin-based quantum computing \cite{Craig-Ramsak,Cho}, and more. The model describes two localized spin-1/2 impurities in an electron gas, coupled by the Ruderman-Kittel-Kasuya-Yosida (RKKY) interaction $\sim J_I$ via their spin exchange with the electrons. In addition to the RKKY coupling the model exhibits a second energy scale, the Kondo temperature $T_K$, below which the electrons may screen the impurity spins. For strong ferromagnetic RKKY interaction, $\mid\!\!J_I\!\!\mid\, \gg T_K$, the impurities form a spin-1 state which does get screened, in exact analogy with the spin-1 two-channel Kondo effect. In contrast, for strong antiferromagnetic RKKY interaction, $J_I \gg T_K$, the impurity spins form a singlet state, killing off the Kondo effect. In the presence of a special electron-hole symmetry \cite{Millis}, {\em or} with the impurities coupled to separate electron reservoirs \cite{Zarand}, the nonuniversal crossover between the two regimes sharpens into a quantum phase transition (QPT) with a non-Fermi liquid quantum critical region.

While much is known about the model, various real-space properties are yet to be uncovered.
A central concept is the {\em Kondo screening cloud}, inferred from the appearance of a characteristic length $\xi_K = \hbar v_F/k_B T_K$ associated with the energy scale $T_K$, with $v_F$ the Fermi velocity \cite{AffleckReview}. The Kondo cloud is invoked to explain how physical quantities at a distance $r$ turn into scaling functions of $r/\xi_K$, rather than depending on $r$ and $\xi_K$ separately \cite{AffleckReview}. However, its nature, structure, and experimental confirmation  has remained controversial, even for the single-impurity Kondo model,  motivating several recent attempts to determine the Kondo length $\xi_K$  accurately \cite{bayat-kondo,Martins}. The case of the TIKM is further compounded since the Kondo regime changes its character as one tunes through zero RKKY coupling (where the model splits into two single-impurity Kondo models) to strong ferromagnetic RKKY coupling (where the physics is instead captured by the spin-1 two-channel Kondo model). What is the signature of this crossover? How does it show up in real space?

In this Letter we address these questions by exploiting and making precise the picture that a screening cloud is built from those electron states that are {\em entangled} with the impurities \cite{bayat-kondo}. This allows us to nonperturbatively uncover: (i) the quantum phase transition between Kondo and RKKY regimes; (ii) the effective impurity-bulk decoupling for antiferromagnetic RKKY coupling; (iii) the true spatial extent of the Kondo cloud; and (iv) the effect of ``Kondo resonance narrowing" for large ferromagnetic RKKY interaction.

\begin{figure} \centering
    \includegraphics[width=8cm,height=7cm,angle=0]{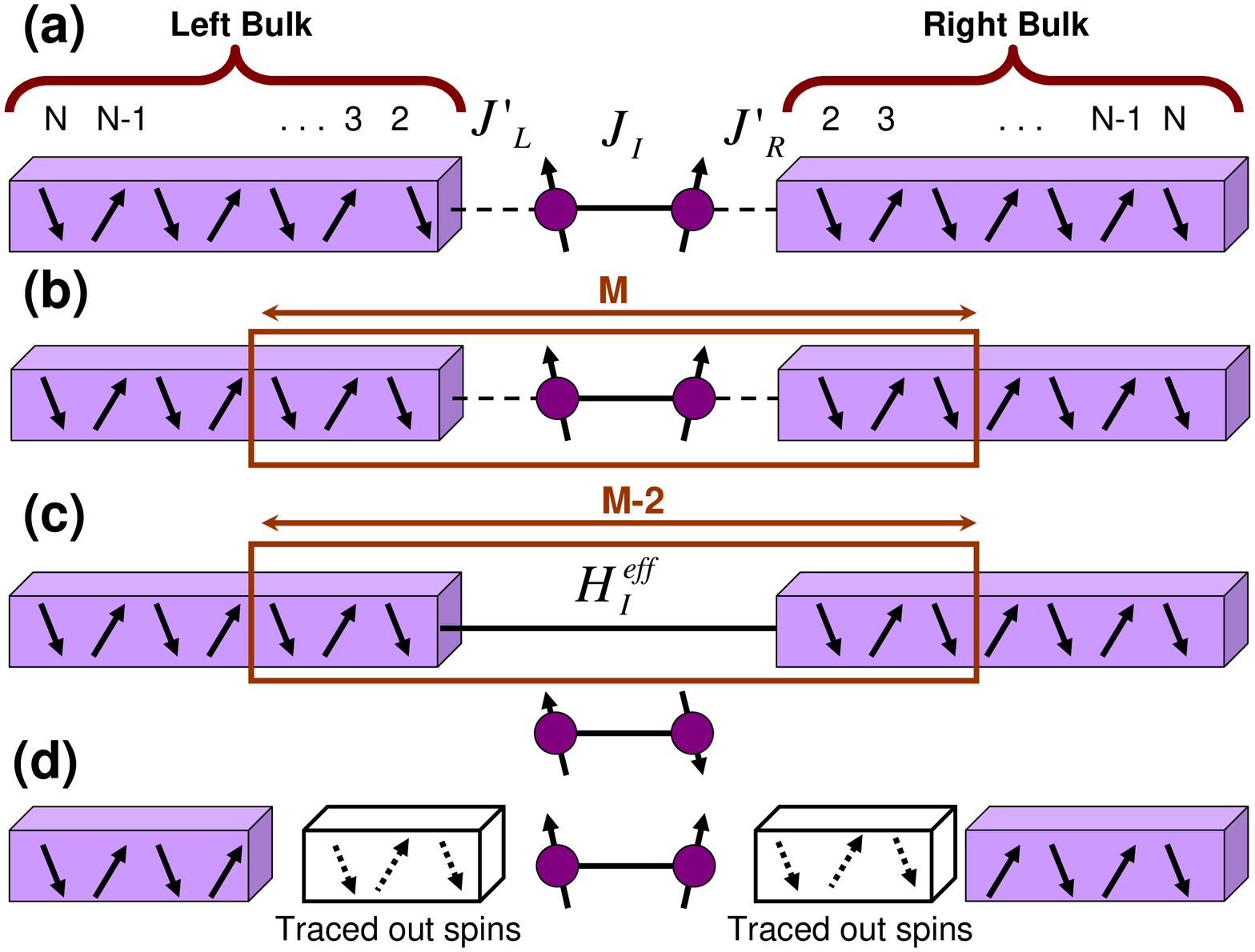}
    \caption{(Color online) (a) The composite system of two Kondo spin chains, each formed by an impurity and a bulk, and coupled together by an RKKY interaction $\sim J_I$ to represent a ``spin-only'' TIKM. (b) The symmetric central block which contains $M$ spins, including impurities. (c) The effective model in which the two impurities are decoupled and instead the bulks are interacting through $H_I^{eff}$, given in Eq. (\ref{H_effective}). (d) Two identical blocks of spins are traced out from the bulks and entanglement between the two impurities and the rest of the system is computed to find the Kondo length.}
     \label{fig1}
\end{figure}

{\em The spin model.-} Exploiting the effective one-dimensionality of the TIKM \cite{ALJ}, we introduce its spin chain emulation by coupling the impurities of two single-impurity Kondo spin chains \cite{LSA} by an RKKY interaction of strength $J_I$, see Fig.~\ref{fig1}(a). The speed-up of numerics achieved by working with  a ``spin-only" version of the TIKM is significant, and enables us to extract entanglement properties via a high-precision Density Matrix Renormalization Group (DMRG) approach. We thus consider the spin Hamiltonian $H= \sum_{k=L,R}H_k + H_I$ where
\begin{eqnarray} \label{Hamiltonian}
H_k &\!=\!& \sum_{j=1}^2 J_j \left( J^{\prime}_k\boldsymbol{\sigma}_1^k \!\cdot \!\boldsymbol{\sigma}_{j+1}^k + \sum_{i=2}^{N_k-j} \boldsymbol{\sigma}_i^k \!\cdot \!\boldsymbol{\sigma}_{i+j}^k \right)
\nonumber \\
H_I &\!=\! &J_I J_1\boldsymbol{\sigma}_1^L \!\cdot \!\boldsymbol{\sigma}_1^R  .
\end{eqnarray}
Here $k=L,R$ labels the left and right chains, with $\boldsymbol{\sigma}_i^k$ the vector of Pauli matrices at site $i$ in chain $k$, and with $J_1$  ($J_2$)  nearest- (next-nearest-) neighbor couplings. Taking $J_1\!=\!1$, $J_2$ must be tuned to a critical value $J_2^c =0.2412$ in order for $H_L$ and $H_R$ to each faithfully represent the spin sector of a single-impurity Kondo model: when $J_2 > J_2^c$ the spin chain enters a gapped dimerized phase, whereas for $J_2 < J_2^c$ a marginal coupling (in the sense of RG) produces logarithmic corrections which pollute numerical data \cite{LSA,sorensen-J1J2}. The positive parameters $J'_L$ and $J'_R$ play the role of antiferromagnetic Kondo couplings. $H_I$, finally, is the RKKY interaction between the impurity spins $\boldsymbol{\sigma}_1^L$ and $\boldsymbol{\sigma}_1^R$, with $J_I$ allowed to take any positive or negative value.  In the DMRG calculations we take $N_L\!=\!N_R\! \equiv N/2$ and assume that $J'_L\! =\!J'_R\!\equiv\!J' $.

{\em Structure of entanglement.-}
For $J_I\!=\!0$, the system decouples into two independent (``spin-only") single-impurity Kondo models where each impurity is screened by its own cloud. For nonzero $J_I$, the ground state attains a more complex structure. To perform a diagnostic we shall use two well known entanglement measures: {\em negativity} \cite{negativity} and {\em von Neumann entropy}. Recall that for a bipartite system $AB$, negativity is defined as ${\cal N}(\rho_{AB})=\sum_i|a_i|-1$, where $a_i$ denote the eigenvalues of the partial transpose of the density matrix $\rho_{AB}$ with respect to one of the two subsets, i.e. $A$ or $B$. The von Neumann entropy, in turn, is given by  $S(\rho_A)=-\mbox{Tr}\rho_A \log_2 \rho_A$, where  $\rho_A=\mbox{Tr}_B \rho_{AB}$ is the reduced density matrix of the subset $A$ when system $B$ is traced out from $\rho_{AB}$. While $S(\rho_A)$ quantifies entanglement only when the total density matrix $\rho_{AB}$ is pure, there is no such restriction for the negativity.

\begin{figure} \centering
    \includegraphics[width=8.5cm,height=4cm,angle=0]{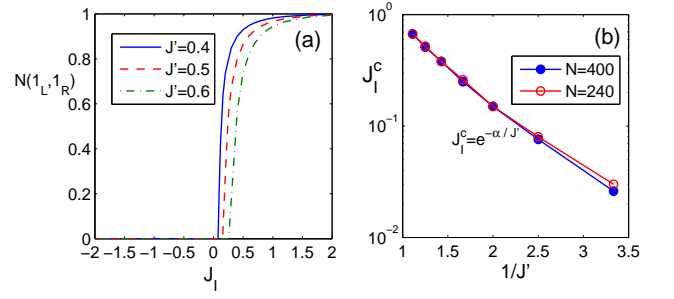}
    \caption{(Color online) (a) Entanglement between two impurities vs. $J_I$ for $N_L=N_R=200$. (b) The coupling $J_I^c$ vs. $1/J'$. }
     \label{fig2}
\end{figure}

Let us first recall that RG studies of the electron-hole symmetric TIKM have identified the fixed-point Hamiltonian for {\em small positive} $J_I$ as that for two independent Kondo impurities \cite{Jones}. However,  unlike for independent spins the ground state expectation value $\la \boldsymbol{\sigma}_1^R \cdot \boldsymbol{\sigma}_1^L\ra$ is nonzero \cite{Jones2}. For $J_I > 2.2 T_K$, no Kondo effect occurs. Still, Kondo correlations persist in this regime, with the two impurity spins locking into a singlet only for very large values of $J_I$. This picture holds also without electron-hole symmetry when the impurity spins couple to separate electron reservoirs with no charge transfer \cite{Zarand}. As this is the case modeled by our Hamiltonian in Eq.  (\ref{Hamiltonian}), we expect a quantum phase transition (QPT) at some value $J_I\!=\!J_I^c$ with $J_I^c$ scaling with $T_K \!\sim\!  e^{-\alpha/J'}$, $\alpha$ being a positive constant \cite{Jones2}. To detect the QPT, we compute the negativity between the impurities as a function of $J_I$. One may take advantage of the SU(2) symmetry of the system and write the reduced density matrix $\rho_{_{1_L,1_R}}$  of the two impurities as a Werner state,
\begin{equation}\label{rho_imp}
 \rho_{_{1_L,1_R}}= P_s |S\ra  \la S|+\frac{1-P_s}{3}\! \sum_{i = 0, \pm 1} |T_i\ra  \la T_i|,
 \end{equation}
where $|S\ra$ is the singlet state, $|T_i\ra$ ($i=0,\pm 1$) are triplets and $P_s$ is the singlet fraction which varies with $J_I$ and $J'$. 
The negativity for a Werner state coincides with its concurrence \cite{Wootters} and can be obtained as ${\cal N}(1_L,1_R) = \mbox{max}\{0,2P_s-1\}$.
The numerical results are depicted in Fig.~\ref{fig2}(a) where the entanglement rises from zero {\em (Kondo regime)} at a point $J_I^c$ (which depends on $J'$) and eventually saturates to unity {\em (local RKKY singlet)}. Moreover, as seen in Fig.~\ref{fig2}(b), the transition point $J_I^c$ indeed scales exponentially with $1/J'$, in agreement with the RG picture from Ref. \cite{Jones2}. The small finite size correction for small $J'$, captured in Fig.~\ref{fig2}(b), reflects the fact that in this parameter regime the extent of Kondo screening cloud approaches the system size.

\begin{figure} \centering
    \includegraphics[width=8.5cm,height=7cm,angle=0]{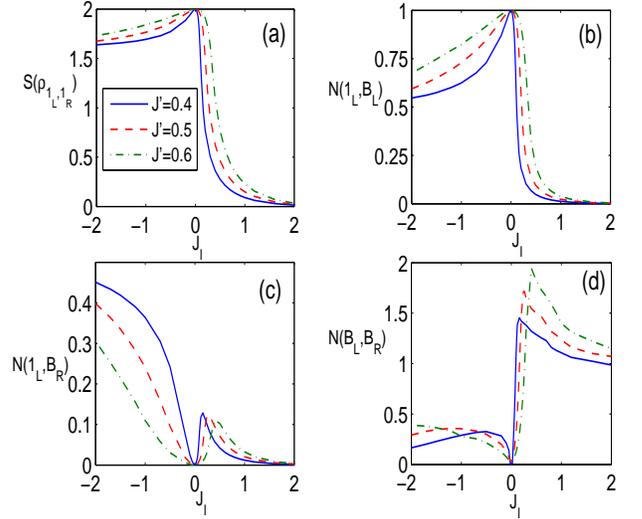}
    \caption{(Color online) (a)  $S(\rho_{_{1_L,1_R}})$ as a function of $J_I$.
    (b) Entanglement between each impurity and its bulk vs. $J_I$. (c) Entanglement between each impurity and its opposite bulk vs. $J_I$.
    (d) Entanglement between the two bulks vs. $J_I$. In all figures $N_L=N_R=200$. }
     \label{fig3}
\end{figure}

By decreasing $J_I$, the singlet fraction $P_s$ decreases monotonically and approaches zero in the limit of large negative $J_I$. It follows that in this limit the two impurity spins effectively behave as a single spin-1 entity, with the two bulks serving as two screening channels. To see this effect one may calculate the von Neumann entropy of $\rho_{_{1_L,1_R}}$,
\begin{equation}\label{S_sL_sR}
    S(\rho_{_{1_L,1_R}})= -P_s \log_2(P_s)-(1-P_s)\log_2 (\frac{1-P_s}{3}).
\end{equation}
As the ground state of the whole system is a pure state, this quantifies the entanglement between the two impurities and the rest of the system. In Fig.~\ref{fig3}(a) we plot $S(\rho_{_{1_L,1_R}})$ as a function of $J_I$ which clearly implies the limiting behavior $\sim \log_2(3)$ (corresponding to $P_s=0$) for $J_I \ll 0$. It is interesting to contrast the Kondo screening behavior for $J_I<0$ with that for $J_I>0$. For $J_I=0$ each impurity is maximally entangled with its neighboring chain independent of the value of $J'$. By additivity of the von Neumann entropy it follows that
 $S(\rho_{_{1_L,1_R}})=2$, as seen also in the DMRG data in Fig.~\ref{fig3}(a).
 Returning to Eq. (\ref{rho_imp}), note that $P_s$ determines the effective impurity spin as a function of $J_I$: For $P_s=0$ (i.e. $J_I \ll 0$) the two impurities behave like a single spin-1 object, while for $P_s=1/4$ (i.e. $J_I=0$) the two impurities are decoupled and each, carrying spin 1/2, gets screened by its own cloud. In the limit $J_I \gg 0$ we have that $P_s = 1$, and the two impurities form a singlet.

It is also instructive to study the entanglement between different constituents of the system.
 In Fig.~\ref{fig3}(b) we display the negativity ${\cal N}(1_L,B_L)$ between the left impurity and the left bulk (by symmetry we have ${\cal N}(1_L,B_L)={\cal N}(1_R,B_R)$). As  $|J_I|$ increases, ${\cal N}(1_L,B_L)$ drops rapidly for $J_I>0$ as one tunes through the QPT where the Kondo screening becomes feeble.  In contrast, for $J_I<0$ the decrease of ${\cal N}(1_L,B_L)$ is slower and approaches a finite value in the limit where the impurity states form a spin-1 state. However, since the left (as well as the right) impurity is now less screened by its own bulk, entanglement monogamy \cite{monogamy} implies that it is entangled also with the opposite bulk, as revealed by Fig.~\ref{fig3}(c).
To display the entanglement between the two bulks for finite $J_I$ we plot the negativity ${\cal N}(B_L,B_R)$ between the left and right chains in Fig.~\ref{fig3}(d), having traced out the impurity states. As Fig.~\ref{fig3}(d) shows, ${\cal N}(B_L,B_R)$ is no longer bounded by unity, reflecting the fact that the bulks contains many spins. Furthermore, due to entanglement monogamy, ${\cal N}(B_L,B_R)$ is larger for $J_I>0$ for which the two impurities tend to decouple from the rest by forming a singlet, in comparison to $J_I<0$ for which the two impurities are entangled with the bulks, thus reducing their ability to get entangled with each other.

\begin{figure} \centering
    \includegraphics[width=8.5cm,height=7cm,angle=0]{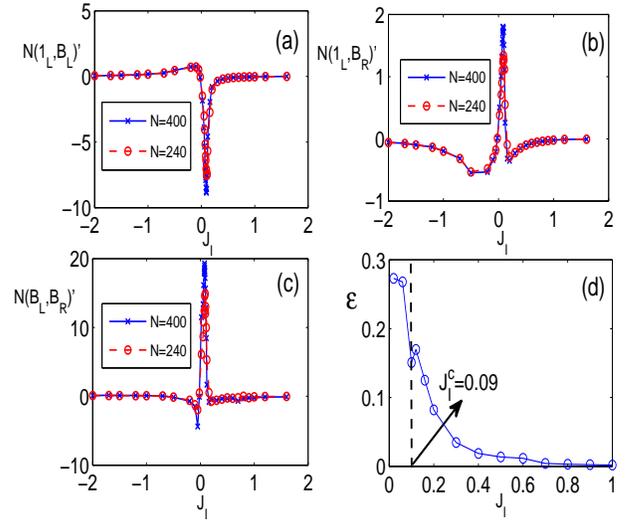}
    \caption{(Color online) The first derivatives with respect to $J_I$ as a function of $J_I$ for $J'=0.4$ and two system sizes $N=240$ and $N=400$: The derivative of (a) ${\cal N}(1_L,B_L)$; (b) ${\cal N}(1_L,B_R)$; (c) ${\cal N}(B_L,B_R)$.  (d) The error parameter $\varepsilon$ vs. $J_I$ for a system of $N=240$ and $J'=0.4$.}
     \label{fig4}
\end{figure}

{\em Quantum phase transition.-} To corroborate that $J_I^c$ is a quantum critical point, we plot the first derivative of the negativities
${\cal N}(1_L,B_L)$, ${\cal N}(1_L,B_R)$ and ${\cal N}(B_L,B_R)$ in Fig.~\ref{fig4}(a)-(c) for two cases, i.e. $N=240$ and $N=400$. The cusps at $J_I^c$, which become increasingly sharper for larger $N$, are finite-size precursors of a divergence in the thermodynamic limit, a hallmark of a second-order QPT \cite{Amico}.
The data in Figs.~\ref{fig4}(a)-(c), together with Fig.~\ref{fig2}(b), provide a highly nontrivial check that our spin chain model
is a faithful (``spin-only'') emulation of the TIKM.

{\em Effective decoupling of impurities.-} To dissect the RKKY regime, $J_I > J_I^c$, we take a central block of $M$ spins which contains both impurities (see Fig.~\ref{fig1}(b))  and compute its von Neumann entropy $S(\rho(M))$.
Deep in the RKKY regime, $J_I\gg J_I^c$,  the two impurities form a singlet and their quantum state becomes pure and decouple from the rest. This can be modeled by having two decoupled impurities in a singlet state together with an effective system of length $N-2$ formed by the left and right bulks as shown in Fig.~\ref{fig1}(c). The effective interaction $H_I^{eff}$ between the two bulks  can be determined by a Schrieffer-Wolff transformation \cite{Schrieffer-Wolff} with the result that
\begin{eqnarray}\label{H_effective}
    H_I^{eff}&\!=\!& \frac{J'^2J_1}{2J_I} \sigma_2^L.\sigma_2^R  + \frac{J'^2 J_2^2}{2J_I J_1} \sigma_3^L.\sigma_3^R \cr
    &\!+\!&\frac{J'^2 J_2}{2J_I} (\sigma_2^L.\sigma_3^L+\sigma_2^R.\sigma_3^R-\sigma_3^L.\sigma_2^R-\sigma_2^L.\sigma_3^R),
\end{eqnarray}
where we have included also the modified boundary interaction {\em within} each chain. Having effectively removed the impurities, the central block can now be considered as that of $M\!-\!2$ sites with density matrix $\rho_{\text{eff}}(M\!-\!2)$. Given this effective model one can compare the von Neumann entropy of the central block, i.e. $S(\rho_{\text{eff}}(M\!-\!2))$, with the one from the original system, i.e. $S(\rho(M))$, and only for $J_I\!\gg \!J_I^c$ do we expect that the two entropies coincide for all $M\!>\!2$. As a test we introduce an error parameter $\mathcal{E}$ which quantifies the average difference between the von Neumann entropy of the central block for the original and effective chain,
\begin{equation}\label{E_L}
    \mathcal{E}(J',J_I)=\frac{1}{N-4} \sum_{M=4}^{N} |S(\rho(M))-S(\rho_{\text{eff}}(M-2))|.
\end{equation}
In Fig.~\ref{fig4}(d) we plot $\mathcal{E}$ as a function of $J_I$ for $J'\!=\!0.4$.
Surprisingly, as revealed by the exponential decay of $\mathcal{E}$, the impurities decouple from the bulk already for intermediate values of $J_I$.

\begin{figure} \centering
    \includegraphics[width=8cm,height=4cm,angle=0]{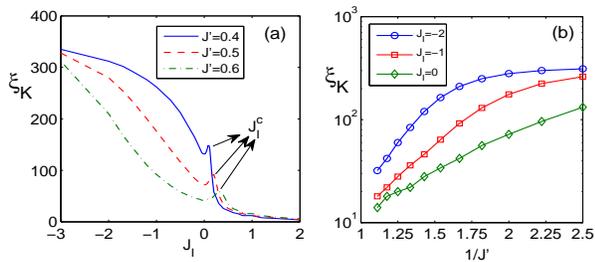}
    \caption{(Color online) (a) Kondo length versus $J_I$. (b) Scaling of $\xi_K$ vs. $1/J'$ for different $J_I$'s. In both figures  $N_L=N_R=200.$ }
     \label{fig5}
\end{figure}
{\em Kondo screening cloud.-} We shall finally address the important issue of the size of the Kondo screening cloud.  Following Ref.  \cite{bayat-kondo} we first trace out two identical blocks of spins, one from each bulk as shown in Fig.~\ref{fig1}(d), and then compute the negativity between the impurities and the rest of the system. The negativity is found to decay exponentially with the number of spins which are traced out. We take the length beyond which the negativity is less than a threshold value, here choosen as 0.01, to define the Kondo length \cite{alternative}, and plot it as a function of $J_I$ in Fig.~\ref{fig5}(a). As seen in the figure, $\xi_K$  increases as one tunes $J_I$ from 0 (with two independent single-impurity Kondo clouds) to large negative values (where the cloud is that of an exactly screened two-channel spin-1 Kondo model). Interestingly, $\xi_K$ increases with small positive values of $J_I$ and takes a small maximum at $J_I^c$. This indicates that the QPT is signaled also in the Kondo screening length at finite system sizes.


\begin{table}
\begin{centering}
\begin{tabular}{|c|c|c|c|c|c|c|c|c|c|}
  \hline
  $J_I$      & -3.00     & -2.50    & -2.00    & -1.50    & -1.00    & -0.50    & 0.00 \\
  \hline
   $\alpha(J_I)$ & 4.1939 & 3.9976 & 3.8861 & 3.1598 & 2.6330 & 2.1262 & 1.7753 \\
  \hline
\end{tabular}
\caption{The values of $\alpha$ for different $J_I$'s in a chain of $N_L=N_R=200$.}
\par\end{centering}
\centering{}\label{table_1}
\end{table}

We have also investigated the scaling of the Kondo length $\xi_K$ as a function of the Kondo coupling $J'$ for different values of $J_I<0$. In Fig.~\ref{fig5}(b) $\xi_K$ is plotted as a function of $1/J'$ in a semi-logarithmic scale. As $\xi_K$ approaches the total length of the system ($N=400$) the curves for negative $J_I$ start bending, indicating that the deviation from linearity is a finite-size effect. One thus infers from Fig.~\ref{fig5}(b) an exponential dependence $\xi_K\sim e^{\alpha/J'}$ in the large-volume limit, {\em with $\alpha$ depending on $J_I$.}
Some values of $\alpha$ are shown in TABLE I. Extrapolation to large negative values of $J_I$ suggests that $\alpha(-\infty)\approx 2.5\alpha(0)$ which is compatible with the results of the analysis carried in Ref.~\cite{NevidomskyyColeman}.
With the Kondo temperature $T_K$ related to $\xi_K$ by $T_K = \hbar v/k_B \xi_K$ \cite{footnote}, our entanglement probe of the Kondo cloud thus shows an exponential suppression of the Kondo temperature with the size of the effective impurity spin (i.e. spin-1 rather than spin-1/2). This phenomenon, known as {\em Kondo resonance narrowing} \cite{NevidomskyyColeman}, was implicitly touched upon in the original work on the TIKM \cite{Jayprakash}. While the problem has been revisited recently in the context of the two-orbital Anderson model \cite{Pruschke} and for magnetic ions with large Hund's coupling  \cite{NevidomskyyColeman}, ours is the first display of the effect extracted from quantum entanglement.

{\em Conclusion.-} We have introduced a spin-chain model representing the two-impurity Kondo model and investigated its properties by applying two complementary entanglement measures borrowed from quantum information theory, {\em negativity} and {\em von Neumann entropy}. This novel approach is conceptually simple and can easily be implemented numerically via a DMRG code. As we have shown in this Letter, it enables one to faithfully recover highly nontrivial features of the two-impurity Kondo model, including the existence of a quantum phase transition at a critical RKKY coupling \cite{Jones}. Importantly, it makes possible, for the first time, a precise probe of how the elusive Kondo cloud depends on the Kondo and RKKY couplings, strikingly showing the effect of Kondo resonance narrowing within a controlled nonperturbative formalism. We expect that our approach can be exploited for generic quantum impurity problems and that it will prove increasingly useful with future applications.


{\em Acknowledgements:-} Discussions with A. Ferraz, A. Hamma, N. Laflorencie, D. Schuricht,  and I. A. Shelykh are gratefully acknowledged.
This work was supported by the Alexander von
Humboldt Foundation and the EU STREPs CORNER, HIP and PICC (AB), Royal Society and
the Wolfson Foundation (SB), and the Swedish Research Council (HJ).

\end{document}